\documentclass[
reprint,
superscriptaddress,
amsmath,amssymb,
aps,
prx,
nofootinbib
]{revtex4-2}

\usepackage{graphicx}
\usepackage{hyperref}
\usepackage{lipsum}
\usepackage{siunitx}
\usepackage{physics}
\usepackage{color}
\usepackage[capitalise]{cleveref}
\usepackage{verbatim}
\usepackage[version=4]{mhchem}
\DeclareSIUnit[]{\cps}{\text{counts/s}}

\newcommand{\EuYSO}{$^{151}$Eu$^{3+}$:Y$_2$Si{O$_5\,$}}
\newcommand{\PrYSO}{Pr$^{3+}$:Y$_2$Si{O$_5\,$}}

\newcommand{\gsi}{\ensuremath{g^{(2)}_{si}}}
\newcommand{\gin}{\ensuremath{g^{(2)}_\text{in}}}
\newcommand{\gout}{\ensuremath{g^{(2)}_\text{out}}}
\newcommand{\stog}{\ket{g}\leftrightarrow\ket{s}}

\newcommand\blfootnote[1]{%
	\begingroup
	\renewcommand\thefootnote{}\footnote{#1}%
	\addtocounter{footnote}{-1}%
	\endgroup
}

\begin{document}
    
\title{Long-lived telecom-heralded single-photon storage in an absorptive spin-rephased quantum memory}

\author{Alberto E. Rodríguez-Moldes}%
\thanks{These authors contributed equally.}
\affiliation{ICFO - Institut de Ciències Fotòniques, The Barcelona Institute of Science and Technology, Castelldefels (Barcelona) 08860, Spain}

\author{Félicien Appas$^{\dagger,\ddagger}$}%
\thanks{These authors contributed equally.}
\affiliation{ICFO - Institut de Ciències Fotòniques, The Barcelona Institute of Science and Technology, Castelldefels (Barcelona) 08860, Spain}
\thanks{These authors contributed equally.}

\author{Jonathan Hänni}%
\affiliation{ICFO - Institut de Ciències Fotòniques, The Barcelona Institute of Science and Technology, Castelldefels (Barcelona) 08860, Spain}

\author{Jelena V. Rakonjac}%
\affiliation{ICFO - Institut de Ciències Fotòniques, The Barcelona Institute of Science and Technology, Castelldefels (Barcelona) 08860, Spain}

\author{Samuele Grandi$^{\S}$}%
\affiliation{ICFO - Institut de Ciències Fotòniques, The Barcelona Institute of Science and Technology, Castelldefels (Barcelona) 08860, Spain}

\author{Hugues de Riedmatten}%
\affiliation{ICFO - Institut de Ciències Fotòniques, The Barcelona Institute of Science and Technology, Castelldefels (Barcelona) 08860, Spain}
\affiliation{ICREA – Institució Catalana de Recerca i Estudis Avançats, 08015 Barcelona, Spain}

\date{\today}
\begin{abstract}
Long-lived storage of single photons under the form of atomic excitations is at the foundation of long-distance entanglement distribution in quantum networks. To mitigate decoherence effects induced by the environment, rephasing of the hyperfine coherences using microwave pulses have been implemented in a variety of single-emitter and ensemble-based solid-state systems. However, the demonstration of storage of single photons in an absorptive quantum memory including such spin rephasing mechanism remains elusive. In this work, we show non-classical storage of telecom-heralded single photons in a \PrYSO rare-earth ion doped crystal quantum memory using the atomic frequency comb (AFC) spin-wave protocol combined with a XY4 spin rephasing sequence. Long-lived AFC photon echoes are first observed in the classical regime for storage times of up to approximately \SI{3}{\milli\second}. We then demonstrate non-classical correlations between heralding photons and stored signal photons generated by a cavity-enhanced parametric photon-pair source for storage times of up to \SI{180}{\micro\second} and with measured cross-correlation values as high as \num{4.6}(4). Together with the capacity of \PrYSO QMs to support highly efficient and multiplexed storage, this result represents a significant step towards scalable long-distance quantum repeater links.
\end{abstract}

\maketitle
\blfootnote{$\dagger$ Corresponding author: felicien.appas@insa-rennes.fr}
\blfootnote{$\ddagger$ Present address: Université de Rennes, INSA Rennes, CNRS, Institut FOTON-UMR 6082, F-35000 Rennes, France}
\blfootnote{$\S$ Present address: Arq Quantum Technologies, Barcelona 08042 Spain}

\begin{figure*}
    \centering
    \includegraphics[width=\linewidth]{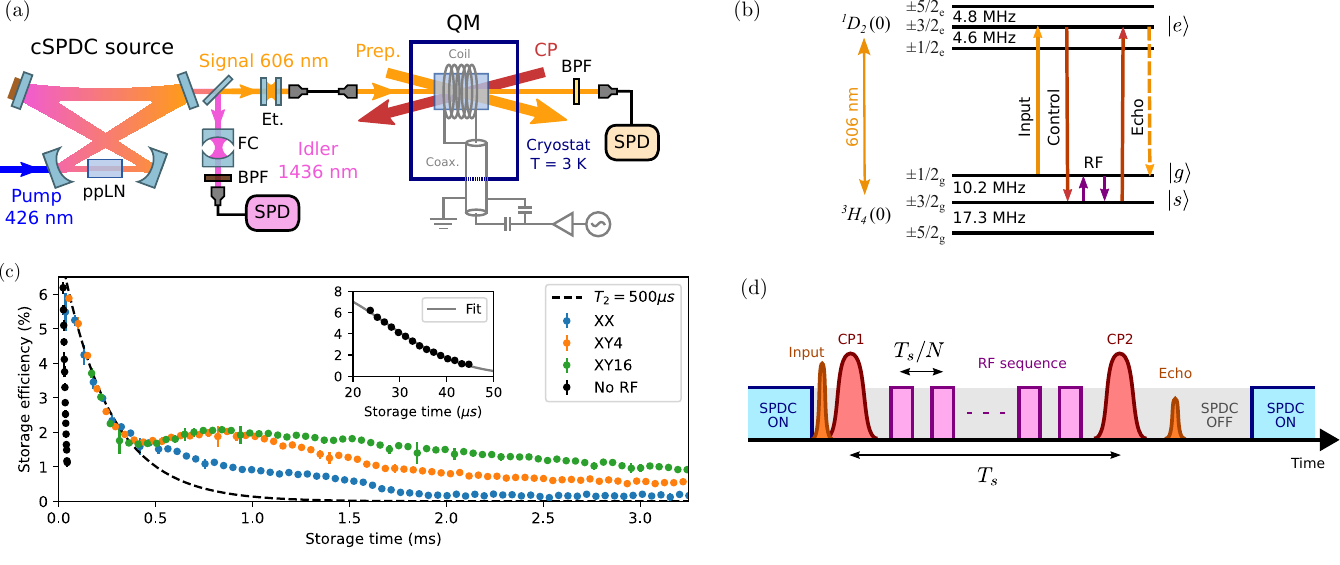}
    \caption{
    (a) Experimental setup. In single photon storage experiments, a second \PrYSO crystal is used for spectral filtering before detection (not shown). ppLN, periodically poled lithium niobate; FC, filter cavity; BPF, bandpass filter; SPD, single photon detector; Et., etalon; Prep., preparation beam; CP, control pulses beam; Coax, coaxial cable.
    (b) Energy levels for the $^{3}H_4(0) \leftrightarrow ^{1}D_2(0)$ transition of \PrYSO showing the optical and hyperfine transitions used in the spin-wave AFC storage protocol.
    (c) Storage efficiency for classical light as a function of total storage time. Black dots: without rf rephasing pulses. Color symbols: single XX, XY4 or XY16 pulse sequence. Black dashed line: calculated decay for XX assuming zero applied magnetic field and a spin coherence time $T_2=\SI{500}{\micro\second}$.
    (d) Schematic representation of the conditional spin-wave AFC storage sequence with spin rephasing ($N$: number of applied rf $\pi$-pulses).}
    \label{fig:fig1}
\end{figure*}

The achievement of large-scale quantum networks is one of the most fascinating pathways of quantum technologies opening the door to exploiting the full potential of quantum computing, sensing and secure communications in a distributed framework~\cite{Kimble2008,Wehner2018}.
Most quantum network architectures rely on the creation of entanglement between stationary matter systems acting as quantum memories (QM) heralded by single photons traveling across long-distance optical channels~\cite{Briegel1998,Duan2001}.
In such a heralded scheme, the stored atomic excitation has to be kept in the QM for at least the total communication round-trip time taken by the heralding photon to travel through the link segment and by the detection signal to travel back to the QM. In typical metropolitan fiber-optical links at telecom wavelengths, this communication time is of the order of \SI{125}{\micro\second} for \SI{25}{\kilo\meter}. However, in a multi-node network, due to losses and the probabilistic nature of the establishment of entanglement, the required storage time for successful entanglement creation over several links is much higher, typically several tens of milliseconds, which poses stringent requirements on the coherence properties of atomic systems to allow for operation in a quantum network scenario. Additionally, to ensure compatibility with fiber architectures and guarantee high entanglement rates at long distance, heralding must be performed at telecom wavelengths.

Long storage times on the hundreds of microseconds to millisecond scale have already been achieved in several types of emissive QMs and exploited in demonstrations of long-distance entanglement distribution. This includes solid-state color centers~\cite{Stolk2024, Knaut2024}, single ions and atoms~\cite{Krutyanskiy2023,vanLeent2022} or laser-cooled atomic ensembles~\cite{Liu2024}.
Yet, the currently limited multiplexing capability and the need to frequency-convert the heralding photon to the telecom band hinder their potential use in a fully-fledged quantum network link.
Another promising system are absorptive rare-earth ion doped crystal (REIC) QMs coupled to narrow-band photon-pair sources~\cite{Simon2007}. Indeed, they benefit from massive multiplexing in several degrees of freedom~\cite{Seri2019,Businger2022,Ortu2022,Teller2025,Li2025a, Ou2025}, high storage efficiency~\cite{Feldmann2025}, telecom heralding at high rate~\cite{Hanni2025} and compatibility with photonic integration~\cite{Zhou2023, Craiciu2021, Rakonjac2022, Liu2025}.
In addition, long-lived storage in the hyperfine nuclear spin state of REICs can be performed by mitigating decoherence effects via spin rephasing sequences applied to the spin transitions~\cite{Viola1999}. Using this technique, ultralong spin coherence times have been observed in multiple REICs species ranging from hundreds of milliseconds to several hours~\cite{Fraval2005,Pascual-Winter2012,Zhong2015,Rancic2018a}. This, in turn, enabled the demonstration of storage of classical light for up to \SI{1}{\hour}~\cite{Ma2021}, single-photon-level pulses for up to \SI{42}{\second}~\cite{Hain2025,Lv2025} as well as non-classical correlations between collective spin excitations and single photons for \SI{1}{\milli\second}, albeit in an emissive scheme and without quantum frequency conversion~\cite{Laplane2017}.

Despite all of these achievements, to date no spin-rephased storage of single photons in an absorptive quantum memory with telecom heralding has been reported. Indeed, previous demonstrations featured reduced storage bandwidth, which implies the use of sub-MHz spectrally narrow photons that are typically difficult to generate using deterministic or parametric sources. In addition, the significant added noise imparted by the multiple rephasing pulses and the reduced count rate when operating with external quantum frequency conversion also make the observation of non-classical storage challenging.

In this Letter, we fill this gap by demonstrating storage of telecom-heralded single photons from a narrowband non-degenerate parametric photon-pair source in a \PrYSO QM using the atomic frequency comb (AFC) protocol and a XY4 spin rephasing sequence. In the classical regime, storage of bright pulses for several millisecond is achieved, the longest reported to date for spin-wave AFC in \PrYSO. Furthermore, we witness non-classical signal/idler correlations at the memory output for storage times of up to \SI{180}{\micro\second}, corresponding to an equivalent fiber-link distance of more than \SI{30}{\kilo\meter}. We emphasize that our scheme does not require a quantum frequency conversion stage and supports the use of moderate photon bandwidth (\SI{2.5}{\mega\hertz}). This result sets the current state of the art for single photon storage in spin-rephased absorptive solid-state quantum memories and represents a milestone towards the accomplishment of long-distance quantum network links.

The experimental setup is sketched in~\cref{fig:fig1}~(a), showing the cavity-enhanced spontaneous parametric down conversion (cSPDC) photon-pair source and the \PrYSO quantum memory located in a 3K closed-cycle cryostat. The relevant energy level scheme is depicted in~\cref{fig:fig1}~(b).
A coil made of copper wire thermally connected to the \SI{3}{\kelvin} stage of the cryostat is wrapped around the memory crystal in order to address the $\stog$ transition at \SI{10.2}{\mega\hertz} via radiofrequency (rf) pulses to perform spin rephasing (see End Matter A).

\begin{figure*}[t]
    \centering
    \includegraphics[width=\linewidth]{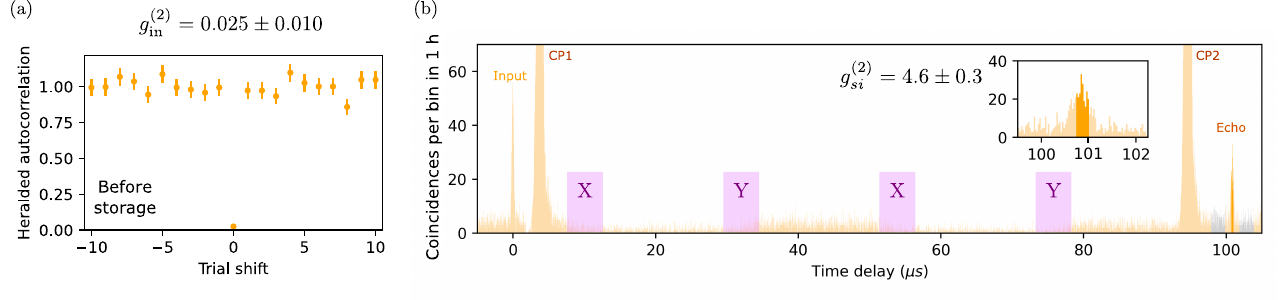}
    \caption{
    (a) Heralded autocorrelation of signal photons without storage. The signal mode goes through a broad transparency window prepared in both the QM and the filter crystal. 
    (b) Time-correlation histogram between the detection of idler and signal photons after storage in the QM with a single XY4 sequence for \SI{100.9}{\micro\second}. The temporal envelope of the rf pulses is depicted in light pink. The inset shows a zoom on the region of the photon echo. Bin size: \SI{20}{\nano\second}}
    \label{fig:fig2}
\end{figure*}

Light storage is performed using the spin-wave AFC protocol (see End Matter B). Extensive descriptions of AFC preparation can be found in Refs.~\cite{Afzelius2009,Jobez2016,Hanni2025}. A comb-shaped absorption profile of periodicity $\Delta$ is prepared into the inhomogeneously broadened $\ket{g}\leftrightarrow\ket{e}$ optical transition by spectral hole burning allowing for storage at a predetermined time of $\tau_\text{AFC} = 1/\Delta$.
To perform on-demand retrieval of the excitation, a pair of bright control pulses (CPs) are used to map the optical excitation in and out of the long-lived hyperfine state $\ket{s}$, as depicted in~\cref{fig:fig1}~(b). By adjusting the time $T_s$ between the two CPs, the recall time $\tau_\text{AFC} + T_S$ can be chosen while the photon is inside the QM, an essential feature for the synchronization of quantum network links.
In this configuration, the storage time is limited by spin dephasing caused by the inhomogeneous broadening of the $\stog$ transition. In \cref{fig:fig1}~(c), we plot in black the storage efficiency as a function of time for a \SI{400}{\nano\second}-long classical input pulse tailored to the spatio-temporal profile of signal photons emitted by the photon-pair source. From the decay, we extract a value of $\SI{15.8(1)}{\kilo\hertz}$ for the spin inhomogeneous broadening, on par with previously reported values in the literature~\cite{Rakonjac2021,Hanni2025}. We observe that spin dephasing prevents the observation of significant photon echoes at storage times larger than \SI{35}{\micro\second} thus severely limiting the use of this method for long-distance quantum networks. 

Fortunately, the effect of spin inhomogeneous broadening can be canceled by inserting between the two CPs a pair of identical $\pi$-pulses at the frequency of the $\stog$ spin transition with a center-to-center temporal separation of $T_S/2$, as sketched in~\cref{fig:fig1}~(d). 
Analogously to a Hahn echo, the spin population is being refocused by the microwave pulses allowing for the observation of AFC echoes at much longer times, as shown in the blue curve of~\cref{fig:fig1}~(c). The storage time under this so-called XX sequence is ultimately limited by the spin coherence time, which in \PrYSO was measured to be $T_2 = \SI{500}{\micro\second}$ at zero magnetic field~\cite{Ham1997}. 
In the case of a $T_2$-limited spin rephasing, the efficiency scales as $\exp(-2T_s/T_2)$ as illustrated by the black dashed line.
However, we see that, in our case, the XX sequence allows for echoes at delays greater the nominal $T_2$. This is likely an indicator of the presence, despite a magnetic shielding layer around the QM, of a small external residual magnetic field inducing Zeeman splitting of the Pr$^{3+}$ and host Yttrium ions hyperfine ground states. This phenomenon translates into an effective decoupling of the rare-earth ion spin from the surrounding Yttrium bath, therefore increasing the spin coherence time~\cite{Pignol2024a}. In this situation, due to the beating of the coherence created between the different spin sublevels, the efficiency decay deviates from a purely exponential law~\cite{Heinze2011,Pignol2024a}. 
In specific field configurations, such as the one observed in our experiment, this can lead to an increase of spin coherence, while in others, it can cause the complete suppression of photon echoes~\cite{Heinze2011,Nicolas2023}. 
We observe a very slight modulation pattern, with a period of approximately $\SI{600}{\micro\second}$, hinting at Zeeman splitting of the Pr$^{3+}$ hyperfine ground state of a few kHz.

More complex sequences with specific phase relationships between the rf $\pi$-pulses allow for an even greater increase of the spin coherence~\cite{Souza2011}. In ~\cref{fig:fig1}~(c), we show the result of storage under a four-pulse XY4 sequence with phases $\{0,\pi/2,0,\pi/2\}$ and a sixteen-pulse XY16 sequence with phases $[\{0,\pi/2,0,\pi/2, \pi/2,0,\pi/2, 0\}]\times 2$. Compared to a simple XX, these sequences benefit from robustness to pulse area errors, meaning the spin coherence can be successfully maintained even in the presence of imperfect rf pulses~\cite{Jobez2015,Zambrini2016}. We see that storage time is indeed notably extended with measurable photon echoes beyond \SI{3}{\milli\second}, which represents an increase of nearly two orders of magnitude compared to previous work on spin-wave AFC storage in \PrYSO without spin rephasing~\cite{Rakonjac2021}.
The efficiency decay with XY4 and XY16 also exhibit non-exponential features which stem from the interplay between the Pr$^{3+}$-Y magnetic dipole coupling under small external magnetic field and the rf pulses position~\cite{Etesse2021}. 
We also emphasize that for storage times above \SI{500}{\micro\second}, the coherence time increases with the number of applied rf pulses, a signature of dynamical decoupling~\cite{Viola1999,Souza2011}. In addition, we point out that the sequence supports on-demand retrieval, similarly to ordinary spin-wave AFC, by delaying the last pulse of the rf sequence while keeping its delay to CP2 fixed.

After characterizing the memory using classical pulses, we perform storage of heralded single photons generated by cSPDC. The source consists of a periodically poled lithiuim niobate crystal inserted inside a doubly-resonant bowtie cavity, as depicted in~\cref{fig:fig1}~(a)~\cite{Fekete2013}. We employ a widely nondegenerate phase-matching with a pump field at \SI{426}{\nano\meter} allowing for the creation of a signal photon resonant with the $\ket{g}\leftrightarrow\ket{e}$ transition of \PrYSO at \SI{606}{\nano\meter} correlated to an idler photon in the telecom band, at \SI{1436}{\nano\meter}. This ensures direct compatibility of the heralding photon with telecom fiber infrastructure without the need to resort to quantum frequency conversion, like in previously demonstrated emissive memory schemes~\cite{Liu2024,Stolk2024,Knaut2024}.
Furthermore, the photons are generated inside the narrow spectral modes of the cavity resulting in a biphoton linewidth of \SI{2.5}{\mega\hertz} that complies with the \SI{4.6}{\mega\hertz} bandwidth of AFC storage in this REIC.
We restrict the generated state to a a single frequency mode by filtering the idler field with a narrowband filter cavity.
Storage is performed in a conditional manner: upon detection of an idler, the SPDC pump is turned off to avoid excess noise in the echo temporal window and the optical CP and rf decoupling pulses are sent to the memory. The storage sequence operates at a repetition rate of 55 attempts per second. A detailed breakdown of all the timing overheads is provided in End Matter C.
After retrieval from the QM, the output field is filtered using a second \PrYSO crystal, located in the same cryostat, into which we prepare a narrow \SI{3}{\mega\hertz}-wide transparency window around the single photon frequency (see End Matter C).

\begin{figure}
    \centering
    \includegraphics[width=0.9\linewidth]{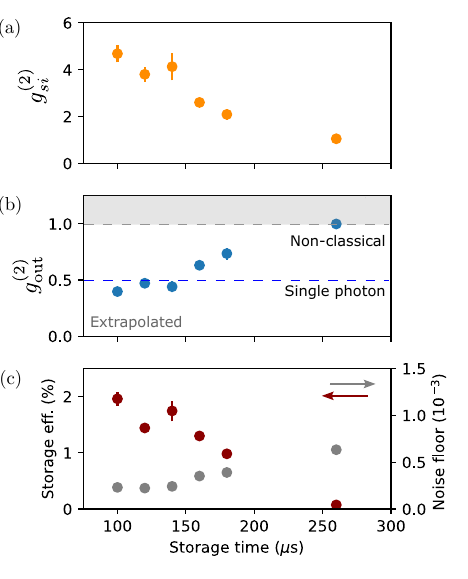}
    \caption{Long-lived storage of heralded single photons with a XY4 sequence. (a) Signal-idler cross correlation $\gsi$, (b) extrapolated output autocorrelation $\gout$ (c) and storage efficiency and noise floor as a function of total storage time. The noise floor here is defined as the probability per herald of detecting a noise photon backpropagated to after the QM.}
  \label{fig:fig3}
\end{figure}

We start by certifying the single-photon nature of the input light by performing a heralded autocorrelation measurement. Here, the signal photons are not stored in the QM but instead are sent through a broad \SI{16}{\mega\hertz}-wide transparency window prepared inside both memory and filter crystals then directed to a fiber 50/50 beam-splitter and detected by superconducting single photon detectors (IDQuantique ID281). The result of the measurement is featured in~\cref{fig:fig2}~(a) where we plot the value of the heralded autocorrelation function of the signal field for different delays between subsequent heralding signals~\cite{Fasel2004}. We see a strong suppression of triple coincidence events at zero delay with a value of $\gin = 0.025\pm0.010$ showing that the cSPDC source emits high-purity heralded single photons.

We then demonstrate the quantum character of the output state by recording temporal correlations between idler and signal detections after spin-wave storage with a single XY4 block (see End Matter D).
The measured time-correlation histogram for a total storage time of \SI{100.9}{\micro\second} is displayed in~\cref{fig:fig2}~(b). We calculate the signal-idler cross correlation as $\gsi = p_{si}/p_sp_i$ where $p_{si}$ is the probability of detecting a coincidence within the \SI{280}{\nano\second} detection window (orange area) while $p_s$ and $p_i$ are the probabilities of single signal and idler clicks. Experimentally, the term $p_sp_i$ corresponds to the noise background of uncorrelated counts and is obtained by integrating coincidence events over two noise windows located before and after the photon detection window (grey areas). The experimental value of the cross-correlation for this storage time is $\gsi =4.6\pm0.3$ which is well above the non-classical bound of 2 for Cauchy-Schwarz inequality violation assuming thermal statistics for the signal and idler fields~\cite{Loudon2006}.
Furthermore, based on the measured values of \gsi and \gin, we extrapolate the heralded autocorrelation at the memory output as~\cite{Heller2022}:
$g^{(2)}_\text{out} = (\gin + 2/s + 1/s^2)/(1+2/s+1/s^2)$,
with $s = \gsi - 1$ the signal-to-noise ratio of the storage in the QM. We infer a value of $g^{(2)}_\text{out} = 0.397 \pm 0.026$ at \SI{100.9}{\micro\second} which is below the threshold of \num{0.5} by \num{3.9} standard deviations, indicating that single photons are retrieved from the memory after storage.

Lastly, we show that spin rephasing allows for successful storage of single photons at longer storage times in the \PrYSO QM. To do so, we repeat the previous measurement for different values of the total storage time. The results are displayed in \cref{fig:fig3}.
As shown in~\cref{fig:fig3}~(a) a value of \gsi above 2 is obtained for up to \SI{180}{\micro\second}, corresponding to an equivalent fiber-link distance of around \SI{36}{\kilo\meter}. In addition, the estimated output autocorrelation $g^{(2)}_\text{out}$, plotted in~\cref{fig:fig3}~(b) is below the single-photon bound of 0.5  for storage times under \SI{150}{\micro\second} and reaches the classical threshold of 1 at \SI{260}{\micro\second}.
This demonstrates the ability of our system to maintain the quantum character of input single photons at storage times that are relevant for metropolitan-scale fiber links and, as such, represents an important step towards the implementation of realistic quantum networks.
 
In the present experiment, the maximum storage time of single photons is limited by the reduced efficiency compared to the classical case together with an increase of the noise background, as can be seen from~\cref{fig:fig3}~(c).
On the one hand, the lower efficiency stems for the fact that the single-photon-storage memory preparation sequence has been optimized to minimize the CP noise in order to reach the highest cross-correlations \gsi which, in general comes at the expense of storage efficiency. Nevertheless, by embedding the QM inside an impedance-matched cavity, single-photon storage efficiencies in spin-wave AFC of up to \SI{18}{\percent} have been demonstrated, therefore providing a favorable solution to this shortcoming~\cite{Feldmann2025}.
On the other hand, the measured increase of noise floor is likely caused by the incapacity of the XY4 sequence to perform perfect rephasing of the spin population in the presence of Zeeman splittings induced by the residual background magnetic field, in spite of the increased spin coherence that was observed in the classical data of~\cref{fig:fig1}~(c)~\cite{Jobez2015,Zambrini2016, Etesse2021}. 
Indeed, in this regime of small fields ($< \SI{1}{\milli\tesla}$) where Zeeman splittings are typically smaller than the Rabi frequency of the $\stog$ transition, the application of an rf pulse leads to a mixing of the Zeeman sublevel populations, as discussed extensively in~\cite{Etesse2021}. This mixing introduces interference between the rephasing contributions from different Zeeman sublevels. As the spacing between rf pulses increases, the dephasing between these quantum trajectories grows, causing destructive interference.
In this situation, efficient spin rephasing requires both a well-chosen magnetic field orientation thanks to the use of a Helmoltz coil assembly and specifically tailored adiabatic rf pulses~\cite{Etesse2021}.
An increasing level of noise with storage time was also observed in early experiments without external field in \EuYSO~\cite{Jobez2015b} but has been successfully overcome in later studies allowing for an improvement of an order of magnitude in storage time at single-photon level~\cite{Ortu2022a}.
This shows that the limitation to the storage time in the present work is not fundamental and can be resolved by implementing similar techniques.

Looking ahead, the use of a well-controlled external magnetic field opens the way to further increasing spin coherence time (see End Matter E). This will be instrumental in the synchronized entanglement swapping operation over several long-distance quantum repeater segments previously limited by the short storage time in the excited state and the lack of on-demand retrieval~\cite{Rakonjac2023,Liu2024a}. Indeed, under moderate magnetic fields ($<$ 10 mT) the predicted \PrYSO single ion spin coherence time can reach up to \SI{5}{\milli\second} without dynamical decoupling~\cite{Pignol2024a} while at higher field amplitudes (up to 100 mT) the QM can be operated in a ZEFOZ configuration allowing for second-scale storage~\cite{Hain2025,Lv2025}. 

In summary, we demonstrated in this Letter long-lived storage of single photons using spin rephasing in an absorptive QM. Storage has been achieved for up to \SI{3}{\milli\second} with classical pulses and up to \SI{180}{\micro\second} with heralded single photons. This result shows the feasibility of storing single photons using spin rephasing sequences, opening the door to ultra-long storage times relevant for long-distance quantum network links. Lastly, the possibility to incorporate high-efficiency and multiplexed storage to our scheme establishes \PrYSO quantum memories as a major candidate for the implementation of scalable quantum network links.

\section*{Acknowledgements}
This project received funding from: Gordon and Betty Moore Foundation (GBMF7446 to H. d. R); Agència de Gestió d'Ajuts Universitaris i de Recerca; Centres de Recerca de Catalunya; Fundació Privada MIR-PUIG; Fundación Cellex; Ministerio de Ciencia e Innovación with funding from European Union NextGeneration funds (MCIN/AEI/10.13039/501100011033, PRTR-C17.I1); Agencia Estatal de Investigación (PID2023-147538OB-I00, Severo Ochoa CEX2024-001490-S); European Union research and innovation program within the Flagship on Quantum Technologies through Horizon Europe project QIA-Phase 1 under grant agreement no. 101102140 and from the Secretariat of Digital Policies of the Government of Catalonia - G.A. GOV/51/2022. F.A. acknowledge funding from the European Union's Horizon 2022 research and innovation program under the Marie Sklodowska-Curie grant agreements No 101104148 (IQARO). S.G. acknowledges funding from ``la Caixa'' Foundation (ID 100010434; fellowship code LCF/BQ/PR23/11980044). J.H. acknowledges funding from the “Secretaria d’Universitats i Recerca del Departament de Recerca i Universitats de la Generalitat de Catalunya” under grant 2024 FI-2 00059, as well as the European Social Fund Plus.

\bibliography{speed.bib}

@article{Jobez2016,
  title = {Towards Highly Multimode Optical Quantum Memory for Quantum Repeaters},
  author = {Jobez, Pierre and Timoney, Nuala and Laplane, Cyril and Etesse, Jean and Ferrier, Alban and Goldner, Philippe and Gisin, Nicolas and Afzelius, Mikael},
  year = 2016,
  month = mar,
  journal = {Physical Review A},
  volume = {93},
  number = {3},
  pages = {032327},
  publisher = {American Physical Society},
  doi = {10.1103/PhysRevA.93.032327},
  urldate = {2023-03-10}
}

@misc{Lv2025,
	title = {Minute-{Scale} {Photonic} {Quantum} {Memory}},
	url = {http://arxiv.org/abs/2511.12537},
	doi = {10.48550/arXiv.2511.12537},
	urldate = {2025-11-23},
	publisher = {arXiv},
	author = {Lv, You-Cai and Zhu, Yu-Jia and Zhou, Zong-Quan and Li, Chuan-Feng and Guo, Guang-Can},
	month = nov,
	year = {2025},
	note = {arXiv:2511.12537 [quant-ph]},
}

@article{Fasel2004,
doi = {10.1088/1367-2630/6/1/163},
url = {https://doi.org/10.1088/1367-2630/6/1/163},
year = {2004},
month = {nov},
publisher = {},
volume = {6},
number = {1},
pages = {163},
author = {Fasel, Sylvain and Alibart, Olivier and Tanzilli, Sébastien and Baldi, Pascal and Beveratos, Alexios and Gisin, Nicolas and Zbinden, Hugo},
title = {High-quality asynchronous heralded single-photon source at telecom wavelength},
journal = {New Journal of Physics}
}

@article{Fekete2013,
  title = {Ultranarrow-{{Band Photon-Pair Source Compatible}} with {{Solid State Quantum Memories}} and {{Telecommunication Networks}}},
  author = {Fekete, Julia and Riel{\"a}nder, Daniel and Cristiani, Matteo and {de Riedmatten}, Hugues},
  year = 2013,
  month = may,
  journal = {Physical Review Letters},
  volume = {110},
  number = {22},
  pages = {220502},
  publisher = {American Physical Society},
  doi = {10.1103/PhysRevLett.110.220502},
  urldate = {2023-04-26}
}

@article{Rakonjac2021,
  title = {Entanglement between a {{Telecom Photon}} and an {{On-Demand Multimode Solid-State Quantum Memory}}},
  author = {Rakonjac, Jelena V. and {Lago-Rivera}, Dario and Seri, Alessandro and Mazzera, Margherita and Grandi, Samuele and {de Riedmatten}, Hugues},
  year = 2021,
  month = nov,
  journal = {Physical Review Letters},
  volume = {127},
  number = {21},
  pages = {210502},
  issn = {0031-9007, 1079-7114},
  doi = {10.1103/PhysRevLett.127.210502},
  urldate = {2023-03-29},
  langid = {english}
}

@article{Seri2019,
  title = {Quantum {{Storage}} of {{Frequency-Multiplexed Heralded Single Photons}}},
  author = {Seri, Alessandro and {Lago-Rivera}, Dario and Lenhard, Andreas and Corrielli, Giacomo and Osellame, Roberto and Mazzera, Margherita and {de Riedmatten}, Hugues},
  year = 2019,
  month = aug,
  journal = {Physical Review Letters},
  volume = {123},
  number = {8},
  pages = {080502},
  publisher = {American Physical Society},
  doi = {10.1103/PhysRevLett.123.080502},
  urldate = {2023-09-04}
}

@misc{Ou2025,
  title = {Multichannel and High Dimensional Integrated Photonic Quantum Memory},
  author = {Ou, Zhong-Wen and Zhu, Tian-Xiang and Liang, Peng-Jun and Hu, Xiao-Min and Zhou, Zong-Quan and Li, Chuang-Feng and Guo, Guang-Can},
  year = 2025,
  month = aug,
  number = {arXiv:2508.19605},
  eprint = {2508.19605},
  primaryclass = {quant-ph},
  publisher = {arXiv},
  doi = {10.48550/arXiv.2508.19605},
  urldate = {2025-11-06},
  archiveprefix = {arXiv},
  langid = {english}
}

@book{Loudon2006,
  title = {The Quantum Theory of Light},
  author = {Loudon, Rodney},
  year = 2006,
  edition = {3. ed., reprint},
  publisher = {Oxford Univ. Press},
  address = {Oxford},
  isbn = {978-0-19-850177-0 978-0-19-850176-3},
  langid = {english}
}

@article{Nicolas2023,
  title = {Coherent Optical-Microwave Interface for Manipulation of Low-Field Electronic Clock Transitions in {{171Yb3}}+:{{Y2SiO5}}},
  shorttitle = {Coherent Optical-Microwave Interface for Manipulation of Low-Field Electronic Clock Transitions in {{171Yb3}}+},
  author = {Nicolas, L. and Businger, M. and Sanchez Mejia, T. and Tiranov, A. and Chaneli{\`e}re, T. and {Lafitte-Houssat}, E. and Ferrier, A. and Goldner, P. and Afzelius, M.},
  year = 2023,
  month = mar,
  journal = {npj Quantum Information},
  volume = {9},
  number = {1},
  pages = {21},
  publisher = {Nature Publishing Group},
  issn = {2056-6387},
  doi = {10.1038/s41534-023-00687-8},
  urldate = {2025-11-06},
  copyright = {2023 The Author(s)},
  langid = {english}
}

@article{Longdell2006,
  title = {Characterization of the Hyperfine Interaction in Europium-Doped Yttrium Orthosilicate and Europium Chloride Hexahydrate},
  author = {Longdell, J. J. and Alexander, A. L. and Sellars, M. J.},
  year = 2006,
  month = nov,
  journal = {Physical Review B},
  volume = {74},
  number = {19},
  pages = {195101},
  publisher = {American Physical Society},
  doi = {10.1103/PhysRevB.74.195101},
  urldate = {2023-06-08}
}

@phdthesis{Jobez2015b,
  title = {{Stockage multimode au niveau quantique pendant une milliseconde}},
  author = {Jobez, Pierre},
  year = 2015,
  doi = {10.13097/ARCHIVE-OUVERTE/UNIGE:83671},
  urldate = {2025-10-31},
  collaborator = {{Afzelius, Mikael (Dir.)} and {Gisin, Nicolas (Dir.)}},
  copyright = {Free access, info:eu-repo/semantics/openAccess},
  langid = {french},
  school = {Universit{\'e} de Gen{\`e}ve}
}

@article{Souza2011,
  title = {Robust Dynamical Decoupling for Quantum Computing and Quantum Memory},
  author = {Souza, Alexandre M. and Alvarez, Gonzalo A. and Suter, Dieter},
  year = 2011,
  month = jun,
  journal = {Physical Review Letters},
  volume = {106},
  number = {24},
  eprint = {1103.4563},
  primaryclass = {cond-mat, physics:physics, physics:quant-ph},
  pages = {240501},
  issn = {0031-9007, 1079-7114},
  doi = {10.1103/PhysRevLett.106.240501},
  urldate = {2022-09-29},
  archiveprefix = {arXiv}
}

@article{Ham1997,
  title = {Frequency-Selective Time-Domain Optical Data Storage by Electromagnetically Induced Transparency in a Rare-Earth-Doped Solid},
  author = {Ham, B. S. and Shahriar, M. S. and Kim, M. K. and Hemmer, P. R.},
  year = 1997,
  month = dec,
  journal = {Optics Letters},
  volume = {22},
  number = {24},
  pages = {1849},
  issn = {0146-9592, 1539-4794},
  doi = {10.1364/OL.22.001849},
  urldate = {2025-10-31},
  copyright = {https://doi.org/10.1364/OA\_License\_v1\#VOR},
  langid = {english}
}

@article{Viola1999,
  title = {Dynamical Decoupling of Open Quantum Systems},
  author = {Viola, Lorenza and Knill, Emanuel and Lloyd, Seth},
  journal = {Phys. Rev. Lett.},
  volume = {82},
  issue = {12},
  pages = {2417--2421},
  numpages = {0},
  year = {1999},
  month = {Mar},
  publisher = {American Physical Society},
  doi = {10.1103/PhysRevLett.82.2417},
  url = {https://link.aps.org/doi/10.1103/PhysRevLett.82.2417}
}

@article{Zhou2023,
  title = {Photonic {{Integrated Quantum Memory}} in {{Rare}}-{{Earth Doped Solids}}},
  author = {Zhou, Zong-Quan and Liu, Chao and Li, Chuan-Feng and Guo, Guang-Can and Oblak, Daniel and Lei, Mi and Faraon, Andrei and Mazzera, Margherita and De Riedmatten, Hugues},
  year = 2023,
  month = oct,
  journal = {Laser \& Photonics Reviews},
  volume = {17},
  number = {10},
  pages = {2300257},
  issn = {1863-8880, 1863-8899},
  doi = {10.1002/lpor.202300257},
  urldate = {2024-12-06},
  langid = {english}
}

@article{Craiciu2021,
  title = {Multifunctional On-Chip Storage at Telecommunication Wavelength for Quantum Networks},
  author = {Craiciu, Ioana and Lei, Mi and Rochman, Jake and Bartholomew, John G. and Faraon, Andrei},
  year = 2021,
  month = jan,
  journal = {Optica},
  volume = {8},
  number = {1},
  pages = {114--121},
  publisher = {Optica Publishing Group},
  issn = {2334-2536},
  doi = {10.1364/OPTICA.412211},
  urldate = {2024-10-24},
  copyright = {{\copyright} 2021 Optical Society of America},
  langid = {english}
}

@article{Knaut2024,
  title = {Entanglement of Nanophotonic Quantum Memory Nodes in a Telecom Network},
  author = {Knaut, C. M. and Suleymanzade, A. and Wei, Y.-C. and Assumpcao, D. R. and Stas, P.-J. and Huan, Y. Q. and Machielse, B. and Knall, E. N. and Sutula, M. and Baranes, G. and Sinclair, N. and {De-Eknamkul}, C. and Levonian, D. S. and Bhaskar, M. K. and Park, H. and Lon{\v c}ar, M. and Lukin, M. D.},
  year = 2024,
  month = may,
  journal = {Nature},
  volume = {629},
  number = {8012},
  pages = {573--578},
  publisher = {Nature Publishing Group},
  issn = {1476-4687},
  doi = {10.1038/s41586-024-07252-z},
  urldate = {2024-09-30},
  copyright = {2024 The Author(s)},
  langid = {english}
}

@article{Laplane2017,
  title = {Multimode and {{Long-Lived Quantum Correlations Between Photons}} and {{Spins}} in a {{Crystal}}},
  author = {Laplane, Cyril and Jobez, Pierre and Etesse, Jean and Gisin, Nicolas and Afzelius, Mikael},
  year = 2017,
  month = may,
  journal = {Physical Review Letters},
  volume = {118},
  number = {21},
  pages = {210501},
  publisher = {American Physical Society},
  doi = {10.1103/PhysRevLett.118.210501},
  urldate = {2025-10-30}
}

@article{Pascual-Winter2012,
  title = {Spin Coherence Lifetime Extension in  {{Tm}}$^{3+}$:{{YAG}} through Dynamical Decoupling},
  shorttitle = {Spin Coherence Lifetime Extension in {{Tm}}\${\textasciicircum}\{3+\}\$},
  author = {{Pascual-Winter}, M. F. and Tongning, R.-C. and Chaneli{\`e}re, T. and Gou{\"e}t, J.-L. Le},
  year = 2012,
  month = nov,
  journal = {Physical Review B},
  volume = {86},
  number = {18},
  eprint = {1208.1622},
  primaryclass = {quant-ph},
  pages = {184301},
  issn = {1098-0121, 1550-235X},
  doi = {10.1103/PhysRevB.86.184301},
  urldate = {2022-09-29},
  archiveprefix = {arXiv}
}

@article{Rancic2018a,
  title = {Coherence Time of over a Second in a Telecom-Compatible Quantum Memory Storage Material},
  author = {Ran{\v c}i{\'c}, Milo{\v s} and Hedges, Morgan P. and Ahlefeldt, Rose L. and Sellars, Matthew J.},
  year = 2018,
  month = jan,
  journal = {Nature Physics},
  volume = {14},
  number = {1},
  pages = {50--54},
  issn = {1745-2473, 1745-2481},
  doi = {10.1038/nphys4254},
  urldate = {2025-10-30},
  langid = {english}
}

@article{Fraval2005,
  title = {Dynamic Decoherence Control of a Solid-State Nuclear Quadrupole Qubit},
  author = {Fraval, E. and Sellars, M. J. and Longdell, J. J.},
  year = 2005,
  month = jul,
  journal = {Physical Review Letters},
  volume = {95},
  number = {3},
  eprint = {quant-ph/0412061},
  pages = {030506},
  issn = {0031-9007, 1079-7114},
  doi = {10.1103/PhysRevLett.95.030506},
  urldate = {2022-09-29},
  archiveprefix = {arXiv}
}

@article{Briegel1998,
  title = {Quantum {{Repeaters}}: {{The Role}} of {{Imperfect Local Operations}} in {{Quantum Communication}}},
  shorttitle = {Quantum {{Repeaters}}},
  author = {Briegel, H.-J. and D{\"u}r, W. and Cirac, J. I. and Zoller, P.},
  year = 1998,
  month = dec,
  journal = {Physical Review Letters},
  volume = {81},
  number = {26},
  pages = {5932--5935},
  issn = {0031-9007, 1079-7114},
  doi = {10.1103/PhysRevLett.81.5932},
  urldate = {2025-10-30},
  copyright = {http://link.aps.org/licenses/aps-default-license},
  langid = {english}
}

@article{Kimble2008,
  title = {The Quantum Internet},
  author = {Kimble, H. J.},
  year = 2008,
  month = jun,
  journal = {Nature},
  volume = {453},
  number = {7198},
  pages = {1023--1030},
  publisher = {Nature Publishing Group},
  issn = {1476-4687},
  doi = {10.1038/nature07127},
  urldate = {2024-10-16},
  copyright = {2008 Springer Nature Limited},
  langid = {english}
}

@article{Wehner2018,
  title = {Quantum Internet: {{A}} Vision for the Road Ahead},
  shorttitle = {Quantum Internet},
  author = {Wehner, Stephanie and Elkouss, David and Hanson, Ronald},
  year = 2018,
  month = oct,
  journal = {Science},
  volume = {362},
  number = {6412},
  pages = {eaam9288},
  publisher = {American Association for the Advancement of Science},
  doi = {10.1126/science.aam9288},
  urldate = {2024-10-16}
}

@article{Pignol2024a,
  title = {Decoherence Induced by Dipole-Dipole Couplings between Atomic Species in Rare Earth Ion Doped {{Y}}$_2${{SiO}}$_5$},
  author = {Pignol, C. and Ortu, A. and Nicolas, L. and D'Auria, V. and Tanzilli, S. and Chaneli{\`e}re, T. and Afzelius, M. and Etesse, J.},
  year = {2024},
  month = dec,
  journal = {Physical Review B},
  volume = {110},
  number = {21},
  pages = {214208},
  publisher = {American Physical Society},
  doi = {10.1103/PhysRevB.110.214208},
  urldate = {2025-07-10}
}

@article{Duan2001,
  title = {Long-Distance Quantum Communication with Atomic Ensembles and Linear Optics},
  author = {Duan, L.-M. and Lukin, M. D. and Cirac, J. I. and Zoller, P.},
  year = {2001},
  month = nov,
  journal = {Nature},
  volume = {414},
  number = {6862},
  pages = {413--418},
  publisher = {Nature Publishing Group},
  issn = {1476-4687},
  doi = {10.1038/35106500},
  urldate = {2024-09-05},
  copyright = {2001 Macmillan Magazines Ltd.},
  langid = {english}
}

@article{Simon2007,
  title = {Quantum {{Repeaters}} with {{Photon Pair Sources}} and {{Multimode Memories}}},
  author = {Simon, Christoph and {de Riedmatten}, Hugues and Afzelius, Mikael and Sangouard, Nicolas and Zbinden, Hugo and Gisin, Nicolas},
  year = {2007},
  month = may,
  journal = {Physical Review Letters},
  volume = {98},
  number = {19},
  pages = {190503},
  publisher = {American Physical Society},
  doi = {10.1103/PhysRevLett.98.190503},
  urldate = {2023-03-20}
}

@article{Zhong2015,
  title = {Optically Addressable Nuclear Spins in a Solid with a Six-Hour Coherence Time},
  author = {Zhong, Manjin and Hedges, Morgan P. and Ahlefeldt, Rose L. and Bartholomew, John G. and Beavan, Sarah E. and Wittig, Sven M. and Longdell, Jevon J. and Sellars, Matthew J.},
  year = {2015},
  month = jan,
  journal = {Nature},
  volume = {517},
  number = {7533},
  pages = {177--180},
  publisher = {Nature Publishing Group},
  issn = {1476-4687},
  doi = {10.1038/nature14025},
  urldate = {2023-01-19},
  langid = {english}
}

@article{Ma2021,
  title = {One-Hour Coherent Optical Storage in an Atomic Frequency Comb Memory},
  author = {Ma, Yu and Ma, You-Zhi and Zhou, Zong-Quan and Li, Chuan-Feng and Guo, Guang-Can},
  year = {2021},
  month = apr,
  journal = {Nature Communications},
  volume = {12},
  number = {1},
  pages = {2381},
  publisher = {Nature Publishing Group},
  issn = {2041-1723},
  doi = {10.1038/s41467-021-22706-y},
  urldate = {2022-09-16},
  copyright = {2021 The Author(s)},
  langid = {english}
}

@article{Ortu2022a,
  title = {Storage of Photonic Time-Bin Qubits for up to 20 Ms in a Rare-Earth Doped Crystal},
  author = {Ortu, Antonio and Holz{\"a}pfel, Adrian and Etesse, Jean and Afzelius, Mikael},
  year = {2022},
  month = dec,
  journal = {npj Quantum Information},
  volume = {8},
  number = {1},
  pages = {29},
  issn = {2056-6387},
  doi = {10.1038/s41534-022-00541-3},
  urldate = {2022-05-17},
  langid = {english}
}

@article{Hain2025,
  title = {Light Storage by Electromagnetically Induced Transparency for One Second at the Level of a Single Photon {{inPr3}}+:{{Y2SiO5}} Prepared with Multiple Frequency Ensembles},
  shorttitle = {Light Storage by Electromagnetically Induced Transparency for One Second at the Level of a Single Photon {{inPr3}}+},
  author = {Hain, Marcel and Stewen, Niklas and Halfmann, Thomas},
  year = {2025},
  journal = {New Journal of Physics},
  issn = {1367-2630},
  doi = {10.1088/1367-2630/adb510},
  urldate = {2025-02-17},
  langid = {english}
}

@article{Stolk2024,
	title = {Metropolitan-scale heralded entanglement of solid-state qubits},
	volume = {10},
	url = {https://www.science.org/doi/10.1126/sciadv.adp6442},
	doi = {10.1126/sciadv.adp6442},
	number = {44},
	urldate = {2024-10-31},
	journal = {Science Advances},
	author = {Stolk, Arian J. and van der Enden, Kian L. and Slater, Marie-Christine and te Raa-Derckx, Ingmar and Botma, Pieter and van Rantwijk, Joris and Biemond, J. J. Benjamin and Hagen, Ronald A. J. and Herfst, Rodolf W. and Koek, Wouter D. and Meskers, Adrianus J. H. and Vollmer, René and van Zwet, Erwin J. and Markham, Matthew and Edmonds, Andrew M. and Geus, J. Fabian and Elsen, Florian and Jungbluth, Bernd and Haefner, Constantin and Tresp, Christoph and Stuhler, Jürgen and Ritter, Stephan and Hanson, Ronald},
	month = oct,
	year = {2024},
	pages = {eadp6442},
}

@article{Liu2024,
  title = {Creation of Memory--Memory Entanglement in a Metropolitan Quantum Network},
  author = {Liu, Jian-Long and Luo, Xi-Yu and Yu, Yong and Wang, Chao-Yang and Wang, Bin and Hu, Yi and Li, Jun and Zheng, Ming-Yang and Yao, Bo and Yan, Zi and Teng, Da and Jiang, Jin-Wei and Liu, Xiao-Bing and Xie, Xiu-Ping and Zhang, Jun and Mao, Qing-He and Jiang, Xiao and Zhang, Qiang and Bao, Xiao-Hui and Pan, Jian-Wei},
  year = {2024},
  month = may,
  journal = {Nature},
  volume = {629},
  number = {8012},
  pages = {579--585},
  publisher = {Nature Publishing Group},
  issn = {1476-4687},
  doi = {10.1038/s41586-024-07308-0},
  urldate = {2024-09-04},
  copyright = {2024 The Author(s), under exclusive licence to Springer Nature Limited},
  langid = {english}
}

@article{Krutyanskiy2023,
  title = {Entanglement of {{Trapped-Ion Qubits Separated}} by 230 {{Meters}}},
  author = {Krutyanskiy, V. and Galli, M. and Krcmarsky, V. and Baier, S. and Fioretto, D. A. and Pu, Y. and Mazloom, A. and Sekatski, P. and Canteri, M. and Teller, M. and Schupp, J. and Bate, J. and Meraner, M. and Sangouard, N. and Lanyon, B. P. and Northup, T. E.},
  year = {2023},
  month = feb,
  journal = {Physical Review Letters},
  volume = {130},
  number = {5},
  pages = {050803},
  publisher = {American Physical Society},
  doi = {10.1103/PhysRevLett.130.050803},
  urldate = {2023-09-26}
}

@article{vanLeent2022,
  title = {Entangling Single Atoms over 33 Km Telecom Fibre},
  author = {{van Leent}, Tim and Bock, Matthias and Fertig, Florian and Garthoff, Robert and Eppelt, Sebastian and Zhou, Yiru and Malik, Pooja and Seubert, Matthias and Bauer, Tobias and Rosenfeld, Wenjamin and Zhang, Wei and Becher, Christoph and Weinfurter, Harald},
  year = {2022},
  month = jul,
  journal = {Nature},
  volume = {607},
  number = {7917},
  pages = {69--73},
  publisher = {Nature Publishing Group},
  issn = {1476-4687},
  doi = {10.1038/s41586-022-04764-4},
  urldate = {2023-01-06},
  copyright = {2022 The Author(s)},
  langid = {english}
}

@article{Teller2025,
  title = {A Solid-State Temporally Multiplexed Quantum Memory Array at the Single-Photon Level},
  author = {Teller, Markus and Plascencia, Susana and Sastre Jachimska, Cristina and Grandi, Samuele and {de Riedmatten}, Hugues},
  year = {2025},
  month = jun,
  journal = {npj Quantum Information},
  volume = {11},
  number = {1},
  pages = {92},
  publisher = {Nature Publishing Group},
  issn = {2056-6387},
  doi = {10.1038/s41534-025-01042-9},
  urldate = {2025-07-10},
  copyright = {2025 The Author(s)},
  langid = {english}
}

@article{Ortu2022,
  title = {Multimode Capacity of Atomic-Frequency Comb Quantum Memories},
  author = {Ortu, Antonio and Rakonjac, Jelena V. and Holz{\"a}pfel, Adrian and Seri, Alessandro and Grandi, Samuele and Mazzera, Margherita and de Riedmatten, Hugues and Afzelius, Mikael},
  year = {2022},
  month = jun,
  journal = {Quantum Science and Technology},
  volume = {7},
  number = {3},
  pages = {035024},
  publisher = {IOP Publishing},
  issn = {2058-9565},
  doi = {10.1088/2058-9565/ac73b0},
  urldate = {2024-04-26},
  langid = {english}
}

@article{Feldmann2025,
  title = {Cavity-{{Enhanced Spin-Wave Solid-State Quantum Memory}}},
  author = {Feldmann, Leo and Wengerowsky, S{\"o}ren and Das, Antariksha and Duranti, Stefano and H{\"a}nni, Jonathan and Grandi, Samuele and {de Riedmatten}, Hugues},
  year = {2025},
  month = sep,
  journal = {Physical Review Letters},
  volume = {135},
  number = {12},
  pages = {120801},
  publisher = {American Physical Society},
  doi = {10.1103/8l9k-12k2},
  urldate = {2025-10-03}
}

@article{Rakonjac2022,
  title = {Storage and Analysis of Light-Matter Entanglement in a Fiber-Integrated System},
  author = {Rakonjac, Jelena V. and Corrielli, Giacomo and {Lago-Rivera}, Dario and Seri, Alessandro and Mazzera, Margherita and Grandi, Samuele and Osellame, Roberto and {de Riedmatten}, Hugues},
  year = {2022},
  month = jul,
  journal = {Science Advances},
  volume = {8},
  number = {27},
  pages = {eabn3919},
  publisher = {American Association for the Advancement of Science},
  doi = {10.1126/sciadv.abn3919},
  urldate = {2023-10-26}
}

@article{Liu2025,
  title = {A Millisecond Integrated Quantum Memory for Photonic Qubits},
  author = {Liu, Yu-Ping and Ou, Zhong-Wen and Zhu, Tian-Xiang and Su, Ming-Xu and Liu, Chao and Han, Yong-Jian and Zhou, Zong-Quan and Li, Chuan-Feng and Guo, Guang-Can},
  year = {2025},
  month = mar,
  journal = {Science Advances},
  volume = {11},
  number = {13},
  pages = {eadu5264},
  publisher = {American Association for the Advancement of Science},
  doi = {10.1126/sciadv.adu5264},
  urldate = {2025-04-22}
}

@article{Li2025a,
  title = {Efficient Storage of Multidimensional Telecom Photons in a Solid-State Quantum Memory},
  author = {Li, Zongfeng and Lei, Yisheng and Kling, Trevor and Hosseini, Mahdi},
  year = {2025},
  month = jun,
  journal = {Optica Quantum},
  volume = {3},
  number = {3},
  pages = {295--302},
  publisher = {Optica Publishing Group},
  issn = {2837-6714},
  doi = {10.1364/OPTICAQ.564321},
  urldate = {2025-10-03},
  copyright = {{\copyright} 2025 Optica Publishing Group},
  langid = {english}
}

@article{Afzelius2009,
  title = {Multimode Quantum Memory Based on Atomic Frequency Combs},
  author = {Afzelius, Mikael and Simon, Christoph and {de Riedmatten}, Hugues and Gisin, Nicolas},
  year = {2009},
  month = may,
  journal = {Physical Review A},
  volume = {79},
  number = {5},
  pages = {052329},
  issn = {1050-2947, 1094-1622},
  doi = {10.1103/PhysRevA.79.052329},
  urldate = {2022-05-19},
  langid = {english}
}

@article{Heinze2011,
  title = {Control of Dark-State Polariton Collapses in a Doped Crystal},
  author = {Heinze, G. and Mieth, S. and Halfmann, T.},
  year = {2011},
  month = jul,
  journal = {Physical Review A},
  volume = {84},
  number = {1},
  pages = {013827},
  issn = {1050-2947, 1094-1622},
  doi = {10.1103/PhysRevA.84.013827},
  urldate = {2025-10-03},
  copyright = {http://link.aps.org/licenses/aps-default-license},
  langid = {english}
}

@article{Heller2022,
  title = {Raman {{Storage}} of {{Quasideterministic Single Photons Generated}} by {{Rydberg Collective Excitations}} in a {{Low-Noise Quantum Memory}}},
  author = {Heller, L. and Lowinski, J. and Theophilo, K. and {Padr{\'o}n-Brito}, A. and De Riedmatten, H.},
  year = {2022},
  month = aug,
  journal = {Physical Review Applied},
  volume = {18},
  number = {2},
  pages = {024036},
  issn = {2331-7019},
  doi = {10.1103/PhysRevApplied.18.024036},
  urldate = {2025-10-08},
  langid = {english}
}

@article{Jobez2015,
  title = {Coherent Spin Control at the Quantum Level in an Ensemble-Based Optical Memory},
  author = {Jobez, Pierre and Laplane, Cyril and Timoney, Nuala and Gisin, Nicolas and Ferrier, Alban and Goldner, Philippe and Afzelius, Mikael},
  journal = {Phys. Rev. Lett.},
  volume = {114},
  issue = {23},
  pages = {230502},
  numpages = {5},
  year = {2015},
  month = {Jun},
  publisher = {American Physical Society},
  doi = {10.1103/PhysRevLett.114.230502},
  url = {https://link.aps.org/doi/10.1103/PhysRevLett.114.230502}
}

@article{Zambrini2016,
author = {Emmanuel Zambrini Cruzeiro and Florian Fröwis and Nuala Timoney and Mikael Afzelius},
title = {Noise in optical quantum memories based on dynamical decoupling of spin states},
journal = {Journal of Modern Optics},
volume = {63},
number = {20},
pages = {2101--2113},
year = {2016},
publisher = {Taylor \& Francis},
doi = {10.1080/09500340.2016.1204472},
URL = { 
    
        https://doi.org/10.1080/09500340.2016.1204472
    
    

},
eprint = { 
    
        https://doi.org/10.1080/09500340.2016.1204472
    
    

}

}

@article{Etesse2021,
  title = {Optical and Spin Manipulation of Non-{{Kramers}} Rare-Earth Ions in a Weak Magnetic Field for Quantum Memory Applications},
  author = {Etesse, J. and Holz{\"a}pfel, A. and Ortu, A. and Afzelius, M.},
  year = {2021},
  month = feb,
  journal = {Physical Review A},
  volume = {103},
  number = {2},
  pages = {022618},
  issn = {2469-9926, 2469-9934},
  doi = {10.1103/PhysRevA.103.022618},
  urldate = {2022-05-16},
  langid = {english}
}

@article{Rakonjac2023,
  title = {Transmission of Light--Matter Entanglement over a Metropolitan Network},
  author = {Rakonjac, Jelena V. and Grandi, Samuele and Wengerowsky, S{\"o}ren and {Lago-Rivera}, Dario and Appas, F{\'e}licien and De Riedmatten, Hugues},
  year = {2023},
  month = dec,
  journal = {Optica Quantum},
  volume = {1},
  number = {2},
  pages = {94},
  issn = {2837-6714},
  doi = {10.1364/OPTICAQ.501048},
  urldate = {2024-12-02},
  langid = {english}
}

@article{Liu2024a,
  title = {Nonlocal Photonic Quantum Gates over 7.0 Km},
  author = {Liu, Xiao and Hu, Xiao-Min and Zhu, Tian-Xiang and Zhang, Chao and Xiao, Yi-Xin and Miao, Jia-Le and Ou, Zhong-Wen and Li, Pei-Yun and Liu, Bi-Heng and Zhou, Zong-Quan and Li, Chuan-Feng and Guo, Guang-Can},
  year = {2024},
  month = oct,
  journal = {Nature Communications},
  volume = {15},
  number = {1},
  pages = {8529},
  publisher = {Nature Publishing Group},
  issn = {2041-1723},
  doi = {10.1038/s41467-024-52912-3},
  urldate = {2024-10-24},
  copyright = {2024 The Author(s)},
  langid = {english}
}

@article{Holzapfel2020,
  title = {Optical Storage for 0.53 s in a Solid-State Atomic Frequency Comb Memory Using Dynamical Decoupling},
  author = {Holz{\"a}pfel, Adrian and Etesse, Jean and Kaczmarek, Krzysztof T and Tiranov, Alexey and Gisin, Nicolas and Afzelius, Mikael},
  year = {2020},
  month = jun,
  journal = {New Journal of Physics},
  volume = {22},
  number = {6},
  pages = {063009},
  issn = {1367-2630},
  doi = {10.1088/1367-2630/ab8aac},
  urldate = {2022-05-16},
  langid = {english}
}

@article{Businger2022,
  title = {Non-Classical Correlations over 1250 Modes between Telecom Photons and 979-Nm Photons Stored in {{171Yb3}}+:{{Y2SiO5}}},
  shorttitle = {Non-Classical Correlations over 1250 Modes between Telecom Photons and 979-Nm Photons Stored in {{171Yb3}}+},
  author = {Businger, M. and Nicolas, L. and Mejia, T. Sanchez and Ferrier, A. and Goldner, P. and Afzelius, Mikael},
  year = {2022},
  month = oct,
  journal = {Nature Communications},
  volume = {13},
  number = {1},
  pages = {6438},
  issn = {2041-1723},
  doi = {10.1038/s41467-022-33929-y},
  urldate = {2024-12-01},
  langid = {english}
}

@article{Hanni2025,
  title = {Heralded {{Entanglement}} of {{On-Demand Spin-Wave Solid-State Quantum Memories}} for {{Multiplexed Quantum Network Links}}},
  author = {H{\"a}nni, Jonathan and {Rodr{\'i}guez-Moldes}, Alberto E. and Appas, F{\'e}licien and Wengerowsky, Soeren and {Lago-Rivera}, Dario and Teller, Markus and Grandi, Samuele and {de Riedmatten}, Hugues},
  year = {2025},
  month = oct,
  journal = {Physical Review X},
  volume = {15},
  number = {4},
  pages = {041003},
  publisher = {American Physical Society},
  doi = {10.1103/wvv1-6lg8},
  urldate = {2025-10-13}
}

\section*{End Matter}
\subsection{Details on the experimental setup}
The QM consists of a \SI{0.05}{\percent}-doped \PrYSO crystal with dimensions 2x3x5 mm cut along the $(D_1,D_2,b)$ crystallographic axes. Input light propagates along the $b$-axis. The crystal is cooled down to a temperature of \SI{3}{\kelvin} using a closed cycle helium cryostat (MyCryoFirm). The crystal is shielded from parasitic external magnetic fields by covering the inner walls of the cryostat radiation shield with high-permeability metallic sheets ($\mu$-metal). Impedance matching of the rf line to the coil surrounding the QM is accomplished using an external circuit of adjustable capacitors. In this configuration we achieve a Rabi frequency of $2\pi \times 19 \text{kHz}$ for the $\stog$ transition at \SI{10.2}{\mega\hertz} with an average rf power of \SI{15}{\watt}. In all measurements featured in this work, we use rf square pulses of \SI{5}{\micro\second} duration without frequency chirp, leading to a FWHM frequency bandwidth of \SI{31.8}{\kilo\hertz}. 

\subsection{Details on the AFC preparation}
Through optical pumping, the atomic population of a single class of ions within the \SI{10}{\giga\hertz} homogeneously broadened absorption line is polarized in the hyperfine state $\ket{g}$ while leaving the state $\ket{s}$ empty. Then, by spectral hole burning, a comb-shaped absorption profile of periodicity $\Delta$ is imprinted onto the $\ket{g}\leftrightarrow\ket{e}$ optical transition over a spectral window of \SI{4.6}{\mega\hertz}, limited by the excited state hyperfine splitting. When an incoming light field is resonant with this transition, a collective excitation is created in the excited state of the atomic ensemble and undergoes dephasing before producing a photon echo in the forward direction after a predetermined storage time of $\tau_\text{AFC} = 1/\Delta$~\cite{Afzelius2009,Jobez2016}.

\subsection{Details on the single-photon storage experimental sequence}
AFC preparation lasts around 400 ms and is synchronized to the cycles of the pulse tube of the cryostat (period of \SI{700}{\milli\second}) to be able to perform storage in the part of the cycle exhibiting the lowest level of mechanical vibrations. Discarding high vibration time windows, the remaining allowed time for single photon storage is 200 ms. For all single-photon measurements, the cSPDC source is operated at a pump power of \SI{1.9}{\milli\watt}, with a heralding effciency of \SI{20}{\percent} and \num{1500} cps raw heralding rate. Details on the source locking scheme can be found in the Appendix of Ref.~\cite{Hanni2025}. By taking into account the source locking (\SI{54}{\percent}) and AFC preparation (\SI{57}{\percent}) duty cycles as well as the \SI{2}{\milli\second} memory dead-time to avoid heating by rf pulses, the rate of storage attempts is on average \num{55} per second. As an illustration, for the data displayed in \cref{fig:fig2}~(b), the measured coincidence rate for a \SI{280}{\nano\second}-long detection window is 0.057 cps and the number of noise photons per herald back-propagated at the memory (noise floor) is \num{0.23e-3}.

This unprecedented low level of noise compared to previous work in \EuYSO can be explained by  the lower rf power required for the rephasing pulses (limiting heating of the sample) as well as a careful optimization of all pulse durations and chirps with respect to noise in single-photon-level measurements. As pointed out in the main text, this emphasis on low noise explains the comparatively low storage efficiency, since sequence optimization was carried out to favor the former figure of merit.

We finally point out that the present experiment is performed conditionally with a single stored temporal mode but can nevertheless be extended to support temporal multiplexing using an unconditional storage sequence similar to the one described in Ref.~\cite{Hanni2025}.

\subsection{Choice of rephasing sequence for single-photon storage}
Aside from XY4, XX and XY8 sequences have been tested in single-photon-level storage experiments but the very high measured NF hindered the observation of echoes at low input photon number. We believe this stems from the imperfect rf pulses which prevent operation both with XX and XY8. Under current conditions, from classical data we estimate the single RF pulse transfer efficiency to be \SI{85}{\percent}. In the case of XY8, the potential improvement provided by the pulse error resilience of the sequence is counterbalanced by the increased amount of noise when doubling the number of rf pulses. In the future, thorough optimization of the rf pulse temporal and spectral properties under controlled B field should allow reaching a near-ideal rf transfer efficiency that may allow working at higher number of rf pulses. 

Furthermore, we estimate that the increase in spin coherence under moderate B fields, or later at a zero-first-order-Zeeman (ZEFOZ) point, will be sufficient to eventually reach relevant timescales for metropolitan quantum communications (millisecond time scale) using only a reduced number of rf pulses~\cite{Pignol2024a,Hain2025} (see End Matter E below). As a consequence, the favorable tradeoff between efficiency and noise of XY4 makes it a good candidate sequence for storage in long-distance quantum network links.

\subsection{Magnetic field regimes for further extension of storage time}
Different regimes of magnetic field amplitude can be leveraged in order to extend the spin coherence time of the \PrYSO quantum memory.
Using low magnetic fields ($<$10 mT), interaction between the Pr$^{3+}$ rare-earth ions and the spin environment of the host crystal can be further mitigated with predicted single-ion spin coherence times of up to \SI{5}{\milli\second} without dynamical decoupling, at optimal field orientations in \PrYSO~\cite{Pignol2024a} and measured single-photon level storage for \SI{20}{\milli\second} in \EuYSO~\cite{Ortu2022a}.
Related to the noise floor increase with storage time discussed in the main text, this field regime should allow suppressing this phenomenon both by eliminating several Y-mediated decoherence mechanisms and by enabling optimal spin population transfer~\cite{Etesse2021}.
At intermediate field magnitudes (10-50 mT), the Zeeman splitting can be large enough such that individual sublevels are adressable individually, allowing for efficient spin rephasing~\cite{Holzapfel2020}.
Eventually, ultra-long storage times can be achieved with magnetic fields of the order of \SI{100}{\milli\tesla} by operating at ZEFOZ configuration and with dynamical decoupling~\cite{Longdell2006}, a technique that enabled the recent demonstration of few-photon-level storage on a second time scale in \PrYSO~\cite{Hain2025} and over \SI{40}{\second} in \EuYSO~\cite{Lv2025}, albeit with ultra-narrow input pulses (\SIrange{3}{7.5}{\micro\second} duration).

\end{document}